# Psoriasis as a consequence of incorporation of beta-streptococci into the microbiocenosis of highly permeable intestines (a pathogenic concept)

**Korotky NG[1], Peslyak MY[2]**

[1] Russian state medical university, Moscow

[2] Holding "Kudits-obraz", Moscow, Web site www.psora.df.ru

**A review of recent investigations into pathogenesis of psoriasis summarizing data on the relationship between the presence beta-streptococci in the organism and the cutaneous immune reaction. Results of the studies on gastrointestinal pathology in psoriatic patients are presented. The authors propose a hypothesis that advocates the primary origin of psoriatic gastrointestinal pathology and secondary nature of skin manifestations. A chronic plaque psoriasis model is developed on the basis of the assumption on the key role of two psorafactors, i.e. hyperpermeability of intestinal walls for certain proteins and incorporation of beta-streptococci in the microbiocenosis of intestinal mucosa. The validity of the model is confirmed by the results of practical clinical work.**

This review is also supposed to establish a link between the abnormalities in GIT (Gastro-Intestinal Tract) and skin manifestations of psoriasis. The results stated in books [1,2,3,4] which were published in the 21st century, served as a cornerstone of the review. Moreover, some earlier books emphasized the organic abnormalities of psoriasis, which result, for example, in a duodenal mucosa, in particular increase in the number of mast cells and eosinophiles or intraepithelial lymphocytes. A number of authors established a link between psoriasis and chronic jejunitis (with a hyperpermeability of small intestine and the smoosmallg of its mucous) [6,7,8]. However, systematic researches in this area have been carried out only in the recent years.

The authors of the study [2] analyze the results of a study of 45 patients having only psoriasis. The study revealed abnormalities in permeability of intestinal walls for carbohydrates and fats correlating with the severity of psoriasis and the duration of disease. The gastroscopy of a part of patients (20 patients) showed diffuse duodenitis in all cases.

The book [1] summarizes the results of the survey of 250 patients having only psoriasis. The research showed the pathology of the upper parts of GIT in all cases. The reduction of epithelial layer correlated with a progressing stage of psoriasis and the duration of disease. The research of biopsies of colons of 20 patients has shown the presence of *"the degenerate-* *dystrophic changes of the epithelial compartment, which first of all affect the absorptive apparatus of colonocytes.*" The authors come to a conclusion that patients with psoriasis face *"the said degenerate-dystrophic changes of cell populations of integumentary and glandular gastrointestinal epithelium with the destruction of important cytoplasmatic organelles of epithelial cell which leads to abnormality of secretion and absorption processes*" [1]. They define these changes as "psoriatic gastrointestinopathy" and assume its secondary nature in relation to skin manifestations of psoriasis.

The study [7] emphasizes the interrelation between psoriasis and intestine microflora. The authors consider microbiocenosis as one of the most probable causes of the increased intestinal permeability.

The role of streptococcal focal infection in causing guttate and plaque psoriasis [3, chapter 5] is well known. At the same time a number of subsequent studies [11,12,13,14] revealed that the basic β-streptococcal antigens (further BS-antigens) provoking and supporting chronic psoriasis are BSP-antigens (β-Streptococci Proteins) – the streptococcal cell and membranous proteins being BS disintegration products. The study [13] revealed that intensive skin immune response takes place against streptococcal proteins of a cell wall with the mass of 20-50 kDA. The said researches conducted by professor Lionel Fry (Great Britain) are supposed to define exactly which BS-antigens provoke and support chronic psoriasis.

The 8th chapter of the study [3] puts forward three possible variants of the pathogenesis of psoriasis. The first is based on the presence of BS-antigens in the skin, the second – on a combination of BS-antigens and cross-reference-reactive self-determinant (for example keratins) and the third – only on the presence of a self-determinant.

The results stated in the study [15] concerning the cases with long duration of psoriasis (more than 10 years) confirm the autoantigenic variant: autoantibodies to fibroblasts (AF) are found in 62% of patients. Meanwhile, this proportion is only 7% at duration of disease of less than one year.

The author of the study [15] and the authors of this study consider the BS-antigenic variant most probable when it comes to the causes and progressing of psoriasis. However, it remains unsettled which factors lead to the presence of BS-antigens in the skin after the elimination of focal BS-infection or without any. The purpose of this study is to form a model, which is supposed to answer this question. In order to form the model it is essential to analyze the results of the researches of intestine microflora and permeability.

# 1. An intestine. The structure and invading mucous and permeability

Inside, the folds of small intestine are covered with villuses containing a network of blood and lymphatic vessels. Enterocytes are the main type of vessels' cells. The enterocytes of the area inverted aside the intestinal lumen are covered with microvillis, which enlarge the absorptive surface of intestine up to 350 sq. m. The life span of enterocytes is usually 3-7 days. The daily volume of absorption is 8.5 liters in small intestine and 0.5 liters in the colon.

Children are born with sterile intestine but colon's microflora starts form while transiting the patrimonial parts and through subsequent contact with mother (the final balance is achieved by the age of 3). The bacterial colonization of GIT proceeds step by step. A healthy person's stomach is practically free of bacteria, the top parts of small intestine containing a comparatively few proportion of them (less than 1000/ml) while the inferior departments and especially the colon contain much more (up to $10^{12}$ /ml of fecal masses).

A smaller part of microbes inhabit the lumen while the majority form colonies on intestinal walls. The colonies cover the epithelium being in symbiosis with intestine and grappling densely with microvillis. The thickness and density of the biofilm, which contributes to parietal digestion, and effective operation of enterocytes and colonocytes rise while approaching to the colon.

The sterility of intestine is harmful for the organism. A normal microecological balance of intestine helps the organism resist intestine infection, contributes to assimilation of nutrients and also increases an individual's life span. Beside pathogenic there are 400 types of admissible and-or necessary microorganisms in the GIT. The question whether there is a proper combination (e.g. for colon) is incorrect.

The proper combination, better to say the proper range of combinations, is determined not only by nutrition, race or age of the individual but also by his own unique features. This means that a wide range of combinations is correct on condition that the proportion of the definite type of bacteria doesn't exceed certain (wide enough) limits.

The chyme transits through small intestine due to its peristaltic activity and then transforms. The substances that the chyme contains (including microorganisms, products of their activity and disintegration) strive to pass through the intestine walls into the blood and-or lymphatic channel because of the difference in osmotic pressure. The enterocytes covered with a biofilm provide the absorption through a number of biochemical processes (depending on what substance is absorbed). Scientists distinguish a) active transport, b) parietal digestion (scission and transformation of the enterocytes' ferments by microorganisms and intestine juice), c) barrier function. A significant malfunction of any of these processes leads to malabsorption syndrome, i.e. to incomplete or, on the contrary, excessive absorption [17].

The microorganisms, which form the biofilm, take an active part in digestion and the absorption of the chyme. One part of the chyme is consumed by the above-mentioned microorganisms while another is transformed into those substances, which are easily absorbed by the organism. In fact the biofilm is a part of the system of digestion. The substances being the products of activity and disintegration of microorganisms forming the biofilm add to the chyme therefore the above-stated facts are also relevant concerning these microorganisms.

For each of materials of chyme the normal rate of intestinal permeability (absorptive function) can be determined as the volume of adsorption of this material per unit of time at its certain quantity, which has acted in an intestine. If the permeability of a certain substance seriously exceeds or is less than the limits it means abnormalities of the absorptive function.

In the article [6] (see section 4) and monograph [2] the results of the research of the psoriatic patients' absorptive function are analyzed. The authors of the last study observe the research of 45 patients with psoriasis. Significant abnormalities of fats and d-xyloses permeability (in 1.5-3 times) were revealed in all cases. The endoscopy of 20 patients revealed chronic diffuse gastroduodenit in all cases.

That means that all patients with psoriasis have serious abnormalities of intestine mucous which seriously affects the permeability. These abnormalities can be both inheritable and caused by gastroenterological diseases.

## 2. Streptococci in skin and intestine

Streptococci (genus Streptococcus) are facultative anaerobes. Streptococci are classified in 17 groups designated by header Latin letters depending on the presence of specific carbohydrates in a cellular wall. There is a separate Braun's classification based on streptococci growth peculiarities at an agar with ram's blood. According to this classification scientists distinguish α-streptococci (partial hemolysis and virescence of the medium, i.e. virescencing), β-streptococci (completely hemolyzing or hemocatheretic) and γ-streptococcuses (producing invisible hemolysis).

Beta-streptococci (the majority of which belong to group A) often cause serious diseases, for instance pharyngitises, cellulitises, erysipelases and streptodermas [18]. The skin manifestations of BS-infections go hand in hand with hyperemia, appearance of phlyctenas, exudation and crusts. Skin integuments possess an innate immunity against streptococcal infections, which consists of various mechanisms of protection against different types of BS depending on the degree and duration of skin infection. The immune response is accompanied by inflammation, increased proliferation of keratinocytes and, as a consequence, appearance of crusts.

In case of focal infection BS-antigens (first of all S. pyogenes) provoke the appearance of temporary rashes (guttate psoriasis) and sometimes permanent ones (chronic plaque psoriasis). This process takes place in a situation when the infection is actually localized far from external skin integuments. The mechanism of this phenomenon has not been completely surveyed yet but there is a broad review of a great deal of data in the study [3, chapter 5].

The majority of streptococci, which belong to group A, are S. pyogenes; therefore these terms are used as synonyms. The studies [3,11,12,13,14] cover the results concerning BS, which belong to group A.

As commensals, streptococci are a part of intestine microflora (their number in excrements exceeds $10^7$ colony-forming units per gramme) [16]. alpha-streptococci are found more frequently but the presence of β-streptococci doesn't cause any intestine infections either. 88 cases of diseases caused by BS are analyzed in the study [20] (BS are divided into serogroups the following way: 43% - A, 27% - B, 4% - C and 26% - G). It is also mentioned that BS are an ordinary part of the microflora of pharynx, skin, intestinal tract and vagina. However, in case of weak immunity BS are prone to pathogenicity (it was not revealed in any case that intestinal tract was the area of BS-pathogenicity). The study [19] provides valuable data concerning the presence of streptococci in different departments of GIT (Table 1).

**Table 1. Number ($10^6$ cell / gramme) of streptococcuses in departments of GIT**

|  | jejunum | ileum | colon | faeces |
|---|---|---|---|---|
| a-Streptococcus (S. mutans, S. salivarius, etc) - oral | 261 | 253 | 1170 | 1691 |
| Streptococcus other, intestinal | 1642 | 127 | 2 | 641 |

The studies [21,22] analyzing the statistics of surgeries on GIT confirm the presence of BS in intestine. This presence was latent before the operation but negatively affected the post-surgery recovering process.

The study [23] is dedicated to a child who had perianal BS-dermatitis that was difficult to diagnose and, as a consequence, guttate psoriasis. Therefore, we can assume that BS-infection of colon's mucous can be brought from perianal area.

Even if permeability is in good condition the products of metabolism and disintegration of microorganisms get into the blood. They are constantly present in blood. Knowing this it is easy to define the exact composition of intestine microflora analyzing blood microbic markers using the method of gas-chromathography combined with mass-spectrometry [24,19]. This method helped to prove that anaerobic cocci (staphylococci, streptococci, enterococci and coryneformed bacteria) form about one-fourth of parietal microflora.

The authors of study [36] have shown, that psoriasis is accompanied by endotoxinemia. Their was a research of antibacterial humoral immunity in relation to a normal and conditional-pathogenic intestinal microflora of patients with psoriasis (32 patients) and a group of healthy persons (120 persons). The level of serumal antibodies was observed in relation to 9 various groups of bacteria, but it exceeded the limit only in relation to S. pyogenes. At patients with psoriasis it on the average more than twice exceeded norm (29.8 mkg/ml against 13.75). At a half of the patients it reached 60 mkg/ml. Traditional treatment has not affected the given parameter in any way.

Knowing the above stated facts we claim that BSP and their fragments are always present in human blood. The degree of their presence, the size and variety of these proteins (and their fragments) depend on two basic factors:
- The permeability of intestine for BSP and their fragments
- The degree of presence of BS-colonies in intestine

These factors are the cornerstone of a new model of the pathogenesis of psoriasis

# 3. The model of pathogenesis of psoriasis

Here (as well as above) and further in the review BSP are β-Streptococci Proteins, which are wall and membranous BS-proteins and their fragments. All last named are disintegration products of BS-colonies. The above-stated facts altogether with the analysis of the known methods and the comparison of particular cases enable us to give the following definition to psoriasis.

**Psoriasis is epidermal hyperproliferation, which is a skin immune system response to the excess of the limit of tolerance to BSP-antigens. The accumulation of BSP-antigens in the skin is caused by their accumulation in blood due to the hyperpermeability of intestinal walls for BSP. BSP are disintegration products of BS-colonies, which are incorporated into the microbiocenosis of intestinal mucous. The hyperpermeability of intestinal walls for BSP is stipulated mainly by genetic factors, but can be also influenced by other, not genetical factors.**

This definition is a brief formulation of the interaction of normal and pathological psorafactors showed on the fig. 1. The novelty of this model in comparison with the one established in book [3] lies in psorafactors 1 and 2 and different formulation of psoriatic cycle. Other factors with similar interrelations are observed in detail and proved in [3, chapter 8]. Let us comment on each psorafactor and put forward the supporting arguments.

**Factor 1. The hyperpermeability of intestinal walls for certain proteins (psorafactor-1)**

This abnormality is determined genetically and-or it can be acquired and can affect both colon and small intestine. This hyperpermeability first of all concerns BSP-antigens, i.e. proteins of cell walls and membranous BS, which appear to be antigens for skin immune system. A number of studies show that hyperpermeability of intestinal walls is present in cases of psoriasis. The treatment of this abnormality is a cornerstone of the effective Pagano's method [4] (see section 4).

In our view, psoriatic gastrointestinopathy [1] is primary related to skin manifestations of psoriasis because one of its consequences is the hyperpermeability of intestinal walls for certain proteins. We assume that the factor of "hyperpermeability of intestinal walls for certain proteins" is hereditable and can be named "psorafactor-1".

The correlation between psoriasis and the abnormalities of the functioning of intestinal walls is irrefutable. For example, Crohn's disease is accompanied by psoriasis in 22% cases [7]. The interrelation between prospective locuses of Crohn's disease and psoriasis is observed in [3, section 1.2.5].

The gastroenterologic diseases, which damage intestinal walls, make psoriasis worse since this damage results in hyperpermeability of intestinal walls. This is proved by the interrelation between the severity of affection of intestinal wall and the severity of psoriasis [2].

In 1999 in Sweden there was an experiment of treatment of psoriasis by a gluten-free diet (see section 4). The exact mechanism of gluten influence on psoriasis is unknown. It is known, however, that gluten (in case of predisposition) can affect the condition of enterocytes (as it happens in cases of Gee's disease). As a consequence, the villuses atrophy which leads to the abnormalities of the absorption of carbohydrates, proteins and fats [17]. It is known that in case of gluten enteropathy the permeability for protein ovalbumin (molecular mass 43 kDa) rises dramatically (60 ng/ml compared to the normal 1-4 ng/ml) [25]. It is possible to assume that this process correlates with the hyperpermeability of intestinal walls for BSP with molecular mass 20-50 kDa [13].

Thus the positive influence of gluten-free diet on patients with psoriasis who also have predisposition to Gee's disease can be revealed. But gluten can cause and support psoriasis by raising the permeability of intestinal walls for BSP not resulting, however, in Gee's disease.

For example, one patient was twice tested (with a year interval) on barrier function (permeability) of small intestine for ovalbumin [25]. The studies of patients having only psoriasis and no serious problems with GIT showed multiple excess of normal limits (14-58 ng/ml compared to normal 1-4 ng/ml). This is an open gate for BSP-antigens…

### Factor 2. The appearance and growth of BS-colonies in the microbiocenosis of intestinal mucous (psorafactor-2)

BS are normal inhabitants (commensals) of intestinal mucous. Significant growth of their colonies takes place due to the enlargement and deepening of the coating of the colon mucous and the possible transition of the colonies to small intestine mucous. As a consequence, this process leads to the increase in the number of BSP-antigens.

This factor of the model of pathogenesis of psoriasis is entirely new. The assumption about the influence of BS-colonies in intestinal microbiocenosis on the development of psoriasis follows naturally out of the models of the pathogenesis of psoriasis stated in [3]. In the author's opinion, BSP-antigens provoke not only the development of guttate psoriasis but also chronic plaque psoriasis. While temporary guttate psoriasis is provoked by temporary focal BS-infection (and in particular cases by temporary perianal BS-dermatitis) then follows a natural question where exactly in the organism BS-colonies can exist constantly without provoking any problems except for maintenance of chronic psoriasis. The answer is simple – the only place where pathogenic BS do not reveal their pathogenicity is intestinal mucous. They can constantly exist there as commensals not attracting the patient's or doctors' attention.

The invasion of intestine mucous starts from the colon. Even then if the permeability for certain proteins is infringed (psorafactor-1) it can result in psoriasis. It is the first stage of psoriasis when small intestine is free from BS-colonies. During this period the situation can be saved by a course of colonics with additives (absorbents, urine, apple Acetum, bifidumbacterine, etc.) which helps to restore the microbiocenosis of the colon, i.e. to destroy BS-colonies completely, to remove the initial cause of psoriasis and also psoriasis itself, at least until another invasion of the colon by BS-colonies.

If no measures are taken to remove BS from the colon and if the motility and state of a patient's intestine are in such condition that there is a significant probability of the transition of BS from the colon to small intestine (constipations, weak ileocaecalis valve) BS will start invading small intestine. This second stage makes the situation worse because it is much more difficult to remove BS-colonies from small intestine since, first of all, small intestine mucous is villiferous and, secondly, colonics do not reach small intestine. The surface of small intestine (including villuses) is much larger than the surface of the colon; therefore, psoriasis develops more seriously. Thus complete purification of the colon is made impossible because there is constant transition of BS-colonies coming from small intestine with chyme.

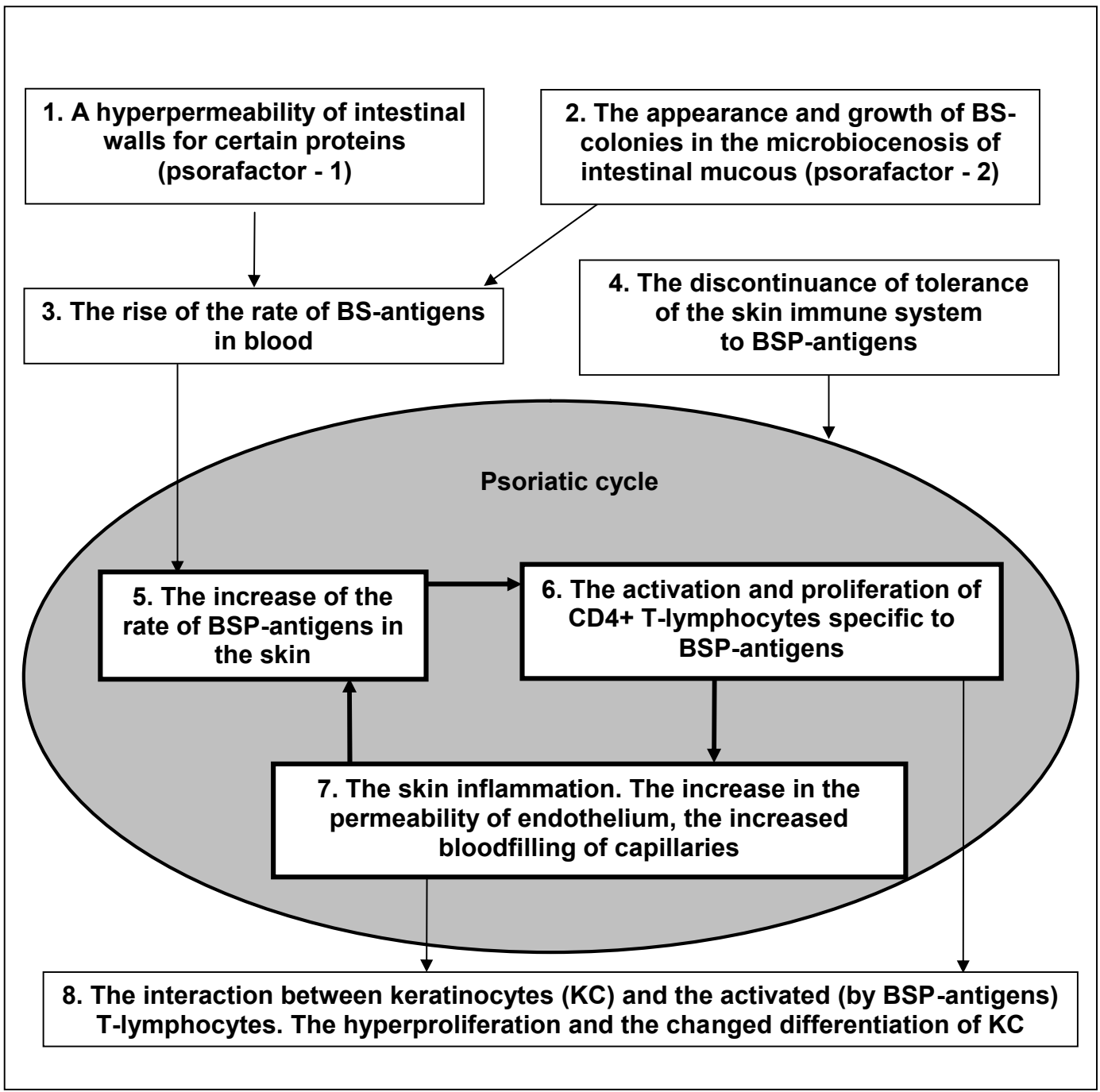

**Fig. 1. Model of pathogenesis of psoriasis. Interaction of factors.**

It is clear from the above-stated facts that it is vital to start the treatment of psoriasis as soon as possible. It becomes evident why the successful results are likely to be achieved at an early stage. Moreover, at an early stage patients can recover entirely. The difficulties of those who started the treatment when the initial cause of psoriasis, the BS-colonization of intestine, had come to the second stage – the invasion of small intestine - are also quite obvious. Unfortunately, the majority of patients belong to the second group.

There is also no surprise in success of those patients who apply special diets and medical starvation. Such methods not only restore the intestinal walls and decrease the permeability for proteins but also undoubtedly influence intestinal microflora. These methods restore intestinal microbiocenosis as a whole especially destroying BS-colonies. Long dry starvation can lead to complete elimination of certain types of microorganisms, for example BS-colonies. Thus this method of treatment of psoriasis is rather effective.

When converting to Pagano's method temporary deterioration, so called Herxheimer reaction [4] can take place, which is well known to doctors since it tends to emerge in the beginning of antibiotic therapy of many bacterial infections. This reaction is provoked by massive destruction of microorganisms accompanied by the increase in the amount of their disintegration products. This phenomenon is a temporary and transient one and normally doesn't require any special treatment. It passes as the pathogenic microorganisms are destroyed by antibiotic therapy.

It is essential to note that in rare cases Herxheimer reaction can lead to more serious consequences than dermal rashes, i.e. to so-called toxic shock that requires medical intervention.

It is possible to assume that the deterioration of skin manifestations connected with converting to Pagano's method is also caused by destruction of BS-colonies existing in intestine mucous (see section 4). Pagano's method results in a change of conditions BS-colonies are accustomed to and in which they multiply which means that they start to perish rapidly due to the lack of nutrients and to the phytotherapy. As a result, their disintegration products get into blood and reach the skin where they force the immune system is made to give an active response.

The fact that applying Pagano's method is not always accompanied by the said reaction means that either a) the gradualness of applying the method determines the gradualness of the destruction of BS-colonies, or b) alive BS-colonies abrupt from intestine's mucous and removal from organism together with chyme and excrements owing to purifying procedures or c) the infringed permeability of intestinal walls is restored faster compared to the tempo of destruction of BS-colonies.

The last assumption is quite probable because the Pagano diet means cutting down on animal proteins, which reduces the load on the intestinal mechanisms of scission, transportation and adsorption of proteins. Suppose that hyperpermeability for BSP-antigens is also connected with the overloading of the said mechanisms. Then its reduction improves their operation and, as a consequence, reduces the permeability of intestinal walls for BSP-antigens.

The application of enterosorbents produces a positive effect on psoriasis (see section 4). In our opinion, its efficiency depends on which amount of disintegration products of BS-colonies is adsorbed (and after that removed from the organism through intestine) before they are splintered to BSP-antigens, which are transported to blood though superpermeable intestinal walls.

Harsh methods (antibiotic therapy, long starvation and so on) altogether with the destruction of BS-colonies can infringe microbiocenosis and lead to disbacteriosis. It is obvious that the answer lies in such a method, which can provide for selective elimination of BS-colonies. That could be achieved by selection of an individual BS-phagolysate and subsequent course of treatment including reception per os and per rectum.

Let's note also the correlation between the following facts: a) people aren't born with chronic psoriasis but it can start at any age and will be always preserved; b) people are born with sterile intestine and the microbiocenosis is formed together with maturing. The changes of intestinal microbiocenosis take place at any age and exist constantly if there is no special treatment or significant change of diet.

We would also like to emphasis the interrelation between early manifestations of psoriasis and artificial suckling of babies. It is well-known that artificial suckling frequently results in a disproportion of intestinal microbiocenosis and decrease of resistance to colonization. Within the limits of the given model it is obvious that that in such conditions BS-colonies have more chances to survive and grow up in the intestinal mucous at an early age of the patient.

In addition, within the limits of this model, it is possible to assume that beside genetic elements family predisposition to psoriasis also includes the factor of living conditions.

For example, if a child who has genetically determined intestinal hyperpermeability for certain proteins lives together with a close relative with psoriasis he is likely to contact BS, the carrier of which is the said close relative. Of course, psoriasis is non-contagious but a person with psoriasis has BS in his intestine (psorafactor-2), which are dangerous for those who have increased intestinal permeability for BSP-antigens (psorafactor-1).

**Factor 3. The rise of the rate of BS-antigens in blood**

This process is caused by factors 1 and 2. This factor is well-known concerning both guttate psoriasis and chronic plaque psoriasis. While the increase of the rate of BSP-antigens at guttate psoriasis is caused by focal BS-infection, the source of BSP-antigens at chronic plaque psoriasis lies in the products of disintegration of BS-colonies as commensals of intestinal microbiocenosis. The products of metabolism and disintegration of microorganisms constantly get into blood but it is namely intestinal hyperpermeability for certain proteins (psorafactor-1) that results in the rise of the rate of BSP-antigens. In the least it reaches the same degree as at focal BS-infection.

In our opinion, a cardinal decrease in of this degree is achieved by such an unsafe method of treatment of heavy psoriasis as hemosorption [27]. Hemosorption provides for the filtration of circulating immune complexes (CIC) and meanmolecular proteins from blood. However, it is not known exactly which antigens the antibodies in CIC are bonded to since the list of filtered meanmolecular proteins remains obscure. In many studies concerning psoriasis the unidentified element filtered by hemosorption is called 'psoragenerative" [28].

Within the limits of the given model it is logical to assume that BSP-antigens are first of all filtered in case of hemosorption, which in the long run leads to the remission of psoriasis. Here it is logical to assume that BSP-antigens are the "psoragenerative".

Another essential effect of hemosorption, i.e. the filtration of CIC, includes the reduction of the load on liver the functions of which are filtration, recycling and detoxification. The decrease of CIC rate in blood also puts an end to the process of their deflection on the endothelium walls of microvessels, which restores their permeability [29].

A number of cleaning procedures (for example the liver cleanse) and medications (for example Essentiale, Essliver Forte) are directly supposed to support the efficient functioning of liver since they support its struggle to recycle BSP-antigens. These procedures and medications enable liver to work so effectively that it becomes capable of refining the blood from the excess of BSP-antigens. As a consequence, a positive result is achieved for the skin but liver gets overloaded which in the long run can affects the patient's health. Besides, the cause of psoriasis is not removed, the result including only temporary decrease of the rate of BSP-antigens in blood.

Haemodesum – a water-salt solution of polyvinylpyrolidone, ions of potassium, sodium, magnesium and chlorine - operates in a similar way. If injected intravenously for a short period (up to 80% is deduced in 4 hours) it binds toxins and deduces them through the renal barrier [30]. The efficiency of Haemodesum in case of serious psoriasis and psoriatic arthritis is well known. However, it remains obscure which psoragenerative it binds. Within the limits of the given model it logical to assume that it binds BSP-antigens.

**Factor 4. The discontinuance of tolerance of the skin immune system to BSP-antigens**

This process takes place because of the excess of the threshold rate and-or due to costimulation[1] since one of the following starting (trigger) events takes place:
- Focal BS-infection
- Streptoderma
- Contagion
- A trauma of epidermis (Koebner effect)

---

- Endocrine deflections
- Reception of certain medicines
- Stress
- Alcohol

The increase of the rate of BSP-antigens in the skin takes place gradually altogether with the growth of BS-colonies in the intestine mucous and the increase in permeability of intestinal walls for certain proteins.

The reaction of skin immune system to BSP-antigens as well as to other BS-antigens is natural. It results in streptoderma (see section 2) and provides the accelerated recovery of skin (hyperproliferation). In case of streptoderma the accelerated reproduction of keratinocytes (KC) provides the accelerated abscission (peeling) of the affected layers of skin, which is an elementary form of protection against infection.

The reaction of skin immune system begins either in case of the excess of the limit of BS-antigens or when a starting event takes place. As a result, lymphocytes specific to BSP-antigens come out of the state of anergy and are activated.

Let's take it into account that any start event is at the same time an event which puts an end to remission and provokes a relapse in case the patient has the experience of illness.

Various starting events differently provoke the beginning (or intensification) of the skin immune system response. For example, trauma itself invokes the inflammatory process, which increases the permeability of endothelium, i.e. directly affects factor 7.

Alcohol affects both intestinal walls increasing their permeability (factor 1) and the permeability of peripheral capillaries (factor 7) thus being a double stimulator. Focal BS-infection and streptoderma provide the growth of the rate of BSP-antigens in blood (factor 3). It is well known that endocrine pathology can affect the intensity and type of the immune response (factor 6). The influence of stress is similar as it actively affects the endocrine state of the organism.

Infections contribute to the increase of the number of the activated (not by BSP-antigens) T-lymphocytes in blood and skin, which leads to the activation of antigen-presenting Langerhans cells (LC). For their part LC express the B7-receptors on their surface, which provides the costimulating, signal while presenting BSP-antigens. Finally, with the help of cytokines it supports the activation of T-lymphocytes specific to BSP-antigens in the skin and-or in the regional lymphonoduses (factor 6).

**Factor 5. The increase of the rate of BSP-antigens in the skin**

This process takes place owing to their migration through endothelium but can be also caused by focal BS-infection.

The skin is one of the eliminative organs, which supplements the operation of liver and contributes to the efforts of liver and kidneys to purify the blood of toxins. In emergency situations (for example in case of intestinal poisoning or contagious disease) the constant peeling of KC provides constant purification of the organism of toxins and various sorts of temporary rashes appear. In fact, chronic psoriasis is constant accelerated blood remission of BSP-antigens (factor 3).

The ability of stationary (attendant) psoriatic plaques to appear mainly on the elbows, knees, feet and hands is well known. The process of transition of BSP-antigens from arteries to skin microvessels is connected with the structure of the arteries, the speed of blood passing through them and the presence of turbulences. The above-mentioned areas have such particularities as flexures and branchings of arteries, which results in greater concentration of BSP-antigens in microvessels and, as a consequence, in their intensive infiltration into the skin.

The positive effect of hydrotherapic procedures such as baths, bathing in cold water, steam baths, an alternating shower is based on active stimulation of the whole system of microvessels which provides for more proportional (though all the surface) infiltration of BSP-antigens into the skin which reduces the intensity of plaques formation.

## Factor 6. The activation and proliferation of CD4+ T-lymphocytes specific to BSP-antigens

This process occurs due to a combination of factors 4 and 5. BSP-antigens are absorbed by Langerhans cells (LC) and processed. Then LC present the processed BSP-antigens to CD4+ T-lymphocytes specific to antigens, which exist in epidermis. At the meantime, other LC deliver BSP-antigens to similar T-lymphocytes of regional lymphonoduses through afferent microvessels. In conditions created by a certain starting event T-lymphocytes are activated to start proliferating.

The activation of epidermal T-lymphocytes results in the release of cytokines, which stimulate the expression of adhesive molecules by endothelial cells of postarterial veinuls.

The proliferation of T-lymphocytes in a regional lymphonodus results in formation of new proliferated T-lymphocytes (including memorial T-lymphocytes), which get into blood-groove through efferent lymphatic vessels. The expression of the homing receptors situated on the proliferated T-lymphocytes provides for their migration to postarterial veinuls on which the adhesive molecules are expressed, i.e. right where LC have activated the epidermal T-lymphocytes.

Their infiltration from the blood-groove to derma takes place straight on the endothelium of the postarterial veinuls owing to the interaction of their receptors and adhesive molecules and then their following migration to epidermis is provoked by hemokines. The increase in number of the activated CD4+ T-lymphocytes (T-helpers) in a certain part of epidermis provides even greater allocation of cytokines, which results in the increase in the permeability of endothelium for leukocytes (neutrophils, macrophages and lymphocytes). Then the inflammatory process starts in the future spot of psoriatic rash. The psoriatic process connected with the increase in number of T-helpers in epidermis is well known [8,30].

It is also well known that Langerhans cells (LC) are ultraviolet (UV) type B sensitive. They are inactivated when affected by UV type B [29, 31]. As a result, the presentation of BSP antigens is prevented as well as the activation of CD4+ T-lymphocytes and the hyperproliferation of KC is slowed down. These facts are the cornerstone of helio- and UV-therapies of psoriasis [4, 30], which result in temporary remission. It is mainly the above-stated facts that determine the seasonal nature of psoriasis since the light day is longer in summer and the weather is warmer, as a result of which people wear less and more transparent clothes which means that skin receives more UV. As a result, LC are inactivated. The situation is quite different in autumn and winter with skin receiving far less UV which means that LC are activated and the relapse of psoriasis occurs.

It is essential to note that these and other traditional therapies [30] suppressing the skin immune response are unable to achieve constant positive results, i.e. to result in complete remission of psoriasis, as they do not influence the psorafactors. Moreover, when the accelerated peeling is stopped while suppressing the skin immune response, factor 3 continues to operate which means that the rate of BSP-antigens in the skin increases even more. And when such therapy comes to an end there is often a relapse of psoriasis the intensity of which depends on the said increased rate of BSP-antigens in the skin.

Having analyzed his enormous practical experience professor Potekaev N.S. came to a conclusion that patients with psoriasis never experience pyoderma (private connection). In our opinion, streptoderma is prevented by constant skin immune response to imaginary presence of BS. Thus the presence of BSP-antigens in the skin provides original constant vaccination against streptodermas.

## Factor 7. The skin inflammation. The increase in the permeability of endothelium, the increased bloodfilling of capillaries

The inflammation is the reaction of the organism, which provides the attraction of leukocytes and the soluble components of plasma to the hearths of infection or tissue damage. Its basic manifestations include the increased bloodfilling of capillaries and their increased permeability for serumal macromolecules and also intensive migration of leukocytes towards the hearth of inflammation through the endothelium of the nearby vessels.

The increased permeability of endothelium for serumal macromolecules means the increased permeability for BSP-antigens. The increased bloodfilling of capillaries results in local increase in the number of BSP-antigens and, as a consequence, the increase of their rate in the skin.

It is also necessary to mark the role of CIC as their deposit on epithelial walls raises their permeability [29, chapter 25]. The increased number of CIC in case of psoriasis is provoked by humoral response to the presence of BSP-antigens in blood as well as other antigens (that is why the infection causing the humoral immune response is one of the starting events) [27,32]. The areas of CIC deposit are likely to define the areas of rashes.

Factors 5,6,7 represent the psoriatic cycle for chronic plaque psoriasis (they are enclosed in an oval on the fig. 1). During the operation of factor 3 this cycle is triggered by one of the costimulating events (factor 4) but continues to operate after factor 3 stops functioning because the started inflammatory process in the skin raises the permeability of endothelium including for BSP-antigens. As a result, inflammation stimulates its causing factor instead of removing it. It means that after the said cycle starts to operate the presence of factor 3 is enough to support it.

The traditional ways of treatment of psoriasis lacerate this cycle affecting factors 6 and 7 at the same time more or less removing (suppressing) the costimulating events (factor 4). However, only the treatment supposed to remove (suppress) at least one of the causes of factor 3, i.e. the treatment affecting factors 1 and 2 will be productive.

Only complete removal of BSP-antigens from blood can in the long run lead to their complete removal from the skin and the ceasing of the inflammatory process.

In case of guttate psoriasis focal BS-infection leads to temporary presence of BSP-antigens in blood. Therefore as BS-infection is fading BSP-antigens are removed from blood after they have succeeded to cause temporary guttate psoriatic rashes.

**Factor 8. The interaction between keratinocytes (KC) and the activated (by BSP-antigens) T-lymphocytes. The hyperproliferation and the changed differentiation of KC**

The interaction between KC and the activated T-lymphocytes takes place after the formation of adhesive molecules on KC. This process is promoted by cytokines produced because of factor 6. The interaction operates through adhesive molecules, other superficial molecules and local production of cytokines. The homology between BSP-antigens and keratins (superficial proteins of KC) can play a stimulating role in particular concerning factor 6 pushing the beginning psoriatic cycle due to cross-reactivity.

This process is a part of chronic inflammation as the accelerated peeling of epidermal cells is the effort of skin immune system to solve the problem of the growing presence of BSP-antigens in the skin. Obviously, this is how skin immune system having used all other opportunities is struggling to get rid of the imaginary presence of BS in the skin. The accelerated peeling is an efficient mechanism against a number of infections when the infecting bacteria are localized in the skin.

In case of chronic psoriasis there is a situation when skin immune system reacts to the increased presence of BSP-antigens erroneously supposing the presence of BS in the skin. The immune system successively turns on the necessary inflammatory mechanisms including (obviously not at once) accelerated peeling. But one of these mechanisms - the increase of permeability of endothelium results in the growing presence of BSP-antigens in the skin. Thus how paradoxical it wouldn't seem the inflammatory process leads to the intensification of its causing factor instead of removing it.

This is how hyperproliferation and the changed differentiation of KC, i.e. chronic psoriasis, are started and supported.

# 4. The practice of treatment

The Pagano method was developed in USA more than 20 years ago and since then it has been used worldwide including Russia. According to Pagano, the cause of psoriasis lies in the abnormality of intestinal barrier function and the penetration of toxins into blood and lymphatic system. Pagano doesn't state in detail what the barrier function is necessary for and which toxins are harmful concerning psoriasis. Besides, he doesn't take into consideration the role of skin immune system. However, his method is mainly supposed to remove psorafactor-1 and often brings positive results.

The Pagano regimen includes special diet, regular purgation, herb teas and the treatment of spine. The purgation is provided by colonics, clysters and unloading fruit diets. Consumption of pure water (1.2-1.6 liters per day) also plays an important role. Pagano diet includes series of restrictions such as the ban on smoking and alcohol, the exclusion of acute, fried, fat and salty products, meat (except for lamb and poultry), shellfishes, mollusks, etc.

**Table 2. Individual values for intestinal permeability. PASI and PSS for study participants**

| Case | Lactulose/mannitol ratio | | PASI | | PSS | |
|---|---|---|---|---|---|---|
| | Before | After | Before | After | Before | After |
| 1 | 0.134* | 0.038 | 7.0 | 4.8 | 7.0 | 6.0 |
| 2 | 0.084* | 0.022 | 30.7 | 18.4 | 14.0 | 5.0 |
| 3 | 0.034 | 0.019 | 14.0 | 0.7 | 21.0 | 3.0 |
| 4 | 0.047 | 0.024 | 2.3 | 0.0 | 7.0 | 1.0 |
| 5 | 0.029 | 0.026 | 37.0 | 19.8 | 24.0 | 12.0 |
| Mean | 0.066* | 0.026 | 18.2 | 8.7 | 14.6 | 5.4 |

* Outside normal range for lactulose/mannitol ratio of 0.01-0.06.

Table 2 displays the results of an experiment, which was held in the USA in 1999 [6]. Five patients, each with a different degree of expression of psoriasis, took part in the experiment. Before converting to Pagano regimen they experienced abnormalities of intestinal permeability (lactulose/mannitolum). This parameter was normalized after six months of following the regimen. The skin condition of the patients also improved (see PASI). In Russia Pagano method also produced successful results, which is confirmed by positive responses of dermatologists and patients.

Another method was applied to treat 30 gluten-dependent patients with psoriasis. First of all, 300 patients with psoriasis were examined in the hospital of Uppsala (Sweden) [33]. In 16% of cases the presence of gliadine antibodies was revealed. It was assumed that a change of diet could improve the skin condition of those patients. An experiment, which involved 30 patients, was held to confirm this hypothesis in 1999. During 3 months the patients followed a gluten-free diet. At the same time they kept applying all the other medications and procedures. In 3 months improvement took place in 73% of cases. 10% of patients showed no changes and deterioration took place in 17% of cases. On average PASI index fell from 5.5 to 3.6. In our opinion, gluten-free diet resulted in partial revival of the intestinal barrier function in relation to BSP-antigens.

The positive effect of applying enterosorbents is well known. The study [8] describes the procedure of applying a medicine called "Sillard" (in Russia manufactured as "Polisorb") consisting of highly disperse silicon dioxide. The treatment starts with a 10-14 day course of "Sillard" therapy. "Sillard" is applied three times a day by doses of 1 g an hour before meal or 1.5 hour after meal. The improvement is observed in the very first days of treatment and is not accompanied by complications. In the study [34] the results of such a therapy (including herb teas) concerning 50 patients are analyzed. Clinical convalescence was observed in 28 cases, significant improvement in 18, improvement took place in 9 cases. In the study [2, chapter 1] the results of successful application of SKNP-1 and SKNP-2 enterosorbents (which are based on activated charcoal) are observed.

In our view, the efficiency of treatment by enterosorbents is connected with the degree of adsorption of BS-colonies disintegration products (including BSP-antigens) before their absorption into blood through intestinal walls.

In the study [35] written in Hungary the authors analyze the hypothesis that the lack of cholic acids (CA) plays its role in the pathogenesis of psoriasis. A group of 551 patients with psoriasis (average PASI =19.1) applied dehydrocholic acid (DA) per os. After the treatment 434 patients (78.8%) became asymptomatic and there was significant improvement in other 117 cases. After the course of treatment the average PASI fell to 2.7. At the meantime, a control group of 249 patients with psoriasis received a traditional therapy after which only 62 patients (24.9%) became asymptomatic. Two years later 319 patients of the first group (57.9%) remained asymptomatic compared to only 15 out of 249 patients of the second group (6%). The application of DA results in temporary increase in volume of CA (including DA), which leads to the reduction of the volume of endotoxins (products of bacterial disintegration) translocated into blood through intestinal walls. The results of the above-mentioned experiment are based on the ability of CA to destroy endotoxins (it isn't specified which exactly).

Suppose these endotoxins are BSP-antigens. However, it is not enough to destroy them for the following prolonged remission because their must be a relapse because BS-colonies producing BSP-antigens have survived. The prolonged asymptomatic remission achieved in the majority of cases gives us an opportunity to assume that the excess of CA destroys BS-colonies. Perhaps, it occurs through CA destroying the adhesive polysaccharide connections of BS-colonies with the mucous-epithelial surface of intestine [16]. Since CA also stimulate the peristalsis the destruction of the connections is followed by the abruption of living BS-colonies from mucous and their subsequent removal from small intestine and organism together with chyme and excrements. Thus the remission can be long and steady and the relapse can occur only owing to new BS-colonization of intestine.

## *References*

1. Nepomnjashchikh GI, Khardikova SA, Ajdagulova SV, Lapii GA, Psoriasis and opisthorchiasis: Morphgenesis of gastrointestinopathy, Moscow, RAMN, 2003, 175 p, Rus.
2. Khardikova SA, Beloborodova EI, Pesterev PN. Psorias, intestinal adsorption. Tomsk, NTL, 2000.120 p, Rus.
3. Baker Barbara S. Recent Advances in PSORIASIS: The Role of the Immune System. ICP Imperial College Press, 2000, 180 p.
4. Pagano John. Healing psoriasis: The natural alternative. The Pagano Organization, 1999; 291 p.
5. Michaelsson G, Kraaz W, Hagforsen E, Pihl-Lundin I, Loof L. Psoriasis patients have highly increased numbers of tryptase-positive mast cells in the duodenal stroma. British Journal of Dermatology. 1997 Jun; 136:866-70.
6. Richards D, Mein E, MakMillin D, Nelson C. Psoriasis case reports, 2000, http://www.meridianinstitute.com/psorias5.html
7. Richards D, Mein E, MakMillin D, Nelson C. Systemic aspects of psoriasis: an integrative model based on intestinal etiology. Integrative Medicine, spring 2000, vol. 2, no. 2, pp. 105-113 (9).
8. Glukhenky BT, Psoriasis, Lechenie i diagnostika, 1,1998, 47, Rus.
9. Galland L. Leaky gut syndromes: breaking the vicious cycle, www.mdheal.org/leakygut.htm
10. Galland L. Power Healing: Use the New Integrated Medicine to Cure Yourself, Random House, 1998, 384 p.

11. Baker BS, Brown DW, Fischetti VA, Ovigne JM, et al. Skin T cell proliferative response to M protein and other cell wall and membrane proteins of group A streptococci in chronic plaque psoriasis, Clin Exp Immunol. 2001 Jun; 124:516-21.
12. Baker BS, Brown D, Fischetti VA, Ovigne JM et al. Stronger proliferative response to membrane versus cell-wall Streptococcal proteins by peripheral blood T cells in chronic plaque psoriasis, Scand J Immunol. 2001 Dec; 54:619-25.
13. Baker BS, Ovigne JM, Fischetti VA, Powles A, Fry L. Selective Response of Dermal Th-1 Cells to 20-50 kDa Streptococcal Cell-Wall Proteins in Chronic Plaque Psoriasis., Scand J Immunol. 2003 Sep; 58:335-41.
14. Brown DW, Baker BS, Ovigne JM, Fischetti VA et al. Non-M protein (s) on the cell wall and membrane of group A streptococci induce (s) IFN-gamma production by dermal CD4 + T cells in psoriasis., Arch Dermatol Res. 2001 Apr; 293:165-70.
15. Markusheva LI, Samsonov VA, Fomina EE, Keda JM, etc., Antibodies to fibroblasts of the skin of the person at patients with various dermatoses, Vestn Dermatol Venerol, 1998, N 2, 31-33, Rus.
16. Shenderov BA. Medical microbial ecology and functional nutrition. In 3 vol, Moscow, Grant, 1998-2001, 286 p; 412 p; 286 p, Rus.
17. Henderson JM, Gastrointestinal pathophysiology. Lippincott, Williams & Wilkins, 1996, 277 p.
18. Medical microbiology. Ed. Pokrovsky VI and Pozdeev OK, Geotar, Medicine, Moscow, 1999, 1184 p, Rus.
19. Parfenov AI, Osipov GA, Bogomolov PO. Intestinal disbacteriosis. Consilium Medicum. 2001; V 3, N 6, Rus.
20. Schugk Jan; Harjola Veli-Pekka; Sivonen Aulikki; Vuopio-Varkila Jaana; Valtonen Matti, A clinical study of beta-haemolytic groups A, B, C and G Streptococcal bacteremia in adults over an 8-year period, Scand J Infect Dis. 1997; 29:233-8.
21. Barnham M. The gut as a source of the haemolytic streptococci causing infection in surgery of the intestinal and biliary tracts. J Infect. 1983 Mar; 6:129-39.
22. Jakobsen J, Andersen JC, Klausen IC, Beta-haemolytic streptococci in acute appendicitis. An ominous finding?, Acta Chir Scand. 1988 Apr; 154:301-3.
23. Herbst RA, Hoch O, Kapp A, Weiss J, Guttate psoriasis triggered by perianal streptococcal dermatitis in a four-year-old boy., J Am Acad Dermatol. 2000 May; 42 (5 Pt 2):885-7.
24. Osipov GA, Parfenov AI, Bogomolov PO, Comparative study of chromatography-mass spectrography of microorganisms' chemical markers in blood and intestinal mucosa bioptats; Ross Gastroenterol Zh. 2001;(1):54-69, Rus.
25. Parfenov AI, Mazo VK, Gmoshinskii IV, Safonova SA., Ekisenina NI, Clinical value of the identification of ovalbumin in blood after a peroral load a dose of proteins of chicken eggs, Ross Gastroenterol Zh, 1999; 2, Rus.
26. Robinson S, My Triumph Over Psoriasis: Curing Disease Without Medication, 1999, 124 p.,
27. Sharapova GI, Korotkii NG, Molodenkov MN, Psoriasis. Moscow. Medicine, 1993, 223 p, Rus.
28. Shilov VN. Psoriasis - solution to the problem. Moscow, 2001., 304 p, Rus.
29. Roitt I, Brostoff J, Male D, Immunology. C.V. Mosby, 2001, 480 p.
30. Hobejsh MM, Moshkalova IA, Sokolovskij EV, Psoriasis. Modern methods of treatment (p.70-128), SOTIS, Saint Petersburg, 1999, 134 p, Rus.
31. Playfair JHL. Immunology at a Glance, Blackwell Publishers, 2000, 96 p.
32. Milevskaya SG, Potapova GV, Characteristics of immune complexes at patients with psoriasis, Vestn dermatol venerol, 1998, N 5, 35-37, Rus.
33. Michaelsson G, Gerden B, Hagforsen E, Nilsson B et al. Psoriatic patients with antibodies to gliadin can be improved by a gluten-free diet. British Journal of Dermatology. 2000 Jan; 142:44-51

34. Trunina TI, The lipid peroxidation condition of psoriatic patients, Lik Sprava, 1998, N 3, с.105-107, Ukr.
35. Gyurcsovics K., Bertok L. Pathophysiology of psoriasis: coping endotoxins with bile acid therapy, Pathophysiology. 2003 Dec; 10:57-61.
36. Garaeva ZS, Safin HA, Kuklin VT, Bildjuk EV, Zinkevich OD, The characteristics of humoral antibacterial immunity at patients with psoriasis, Kazan medical magazine, 2001, N 5, 359-61, Rus.